\documentclass[aps,preprint,superscriptaddress,amsmath,amssymb,nofootinbib]{revtex4}
\usepackage{graphicx,color}
\usepackage{subfigure,bbold,slashed}

\begin{document}

{\begin{flushright}{KIAS-P16041}
\end{flushright}}

\title{ Phenomenology of an $SU(2)_1 \times SU(2)_2 \times U(1)_Y$ model at the LHC}

\author{ Chuan-Hung Chen \footnote{Email: physchen@mail.ncku.edu.tw} }
\affiliation{Department of Physics, National Cheng-Kung University, Tainan 70101, Taiwan }

\author{ Takaaki Nomura \footnote{Email: nomura@kias.re.kr }}
\affiliation{School of Physics, Korea Institute for Advanced Study, Seoul 02455, Republic of Korea}

\date{\today}

\begin{abstract}

 We  investigate the implications of a minimal  $SU(2)$ gauge symmetry extension of the standard model at the LHC.  To achieve the spontaneous symmetry breaking, a heavy Higgs doublet of the $SU(2)$ is introduced.  To obtain an anomaly free model and the decays of new charged gauge bosons, we include a vector-like quark doublet.  We  also employ a real scalar boson to dictate  the heavy Higgs production via the gluon-gluon fusion processes. 
 It is found that the new gauge coupling and the masses of new gauge bosons can be strictly bounded by the electroweak $\rho$-parameter and   dilepton resonance experiments at the LHC.  It is found that due to the new charged gauge boson enhancement,  the  cross sections for  a heavy scalar boson to  diphoton channel measured by ATLAS and CMS  can be easily satisfied when the values of Yukawa couplings are properly taken. Furthermore, by adopting event simulation, we find that   the significance of $pp\to (\gamma \gamma)_H+{\rm jet}$, where the diphoton is from the heavy Higgs decay, can be over $4\sigma$ when the luminosity is above 60 fb$^{-1}$. 
 
\end{abstract}

\maketitle

\section{Introduction}

The Large Hadron Collider (LHC) can not only  test the standard model (SM) but also probe the physics beyond the SM, enabling the exploration of new physics.  
 Some potential events indicating the effects of new physics have indeed been reported by the  ATLAS and CMS experiments. For instance,  diboson resonance  at around 2 TeV was shown by ATLAS~\cite{Aad:2015owa} and CMS~\cite{Khachatryan:2014hpa}; an unexpectedly large branching ratio (BR) for  $h\to \mu \tau$   was given by CMS~\cite{Khachatryan:2015kon}.
The search for  new resonances has been  performed  by CMS and ATLAS  in the dijet decays at the center-of-energy of $\sqrt{s}=13$ TeV ~\cite{Khachatryan:2015dcf,ATLAS:2015nsi}  and in the dilepton  channels~\cite{Chatrchyan:2012oaa,Aad:2014cka,ATLAS:dilepton}. Although no significant excess has been observed, the  data can give strict limits on the mass of resonance and its couplings to the SM particles. 

 Moreover, a hint of a new resonance with a mass of around 750 GeV  in the diphoton spectrum was reported by  the ATLAS~\cite{ATLAS-CONF-2015-081,Aaboud:2016tru} and CMS~\cite{CMS:2015dxe,Khachatryan:2016hje} experiments.
 Inspired by the measurements, the diphoton excess issue was broadly discussed~\cite{Harigaya:2015ezk,Backovic:2015fnp,Angelescu:2015uiz,Nakai:2015ptz,Buttazzo:2015txu,DiChiara:2015vdm,Knapen:2015dap,Pilaftsis:2015ycr,Franceschini:2015kwy,Ellis:2015oso,Gupta:2015zzs,Kobakhidze:2015ldh,Falkowski:2015swt,Benbrik:2015fyz,Wang:2015kuj,Dev:2015isx,Allanach:2015ixl,Wang:2015omi,Chiang:2015tqz,Martinez:2015kmn,Chao:2015nsm,Chang:2015bzc,Feng:2015wil,Boucenna:2015pav,Hernandez:2015ywg,Dey:2015bur,Pelaggi:2015knk,deBlas:2015hlv,Huang:2015rkj,Patel:2015ulo,Das:2015enc,Cao:2015scs,Dev:2015vjd,Jiang:2015oms,Dasgupta:2015pbr,Ko:2016lai,Karozas:2016hcp,Modak:2016ung,Dutta:2016jqn,Deppisch:2016scs,Berlin:2016hqw,Borah:2016uoi,Ko:2016wce,Hati:2016thk,Yu:2016lof,Dorsner:2016ypw,Faraggi:2016xnm,Aydemir:2016qqj,Staub:2016dxq,Ko:2016sxg,Ren:2016gyg,Lazarides:2016ofd,Aydemir:2016xtj,Huong:2016kpa,Leontaris:2016wsy,Kownacki:2016hpm,Nilles:2016bjl,Duerr:2016eme,Takahashi:2016iph,Dutta:2016ach,Davoudiasl:2016yfa,Cao:2015apa}.  However, a clearer signature of the diphoton resonance  is not confirmed by the updating data of ATLAS~\cite{ATLAS:2016eeo} and CMS~\cite{Khachatryan:2016yec}, and the significance of the resonance is somewhat diminished. Nevertheless,  it is still  a good channel to probe a new resonance through the diphoton decay. 

 Since several observed phenomena have not been resolved yet, such as the origin of neutrino masses, dark matter, and anomalous muon $g-2$, it is believed that the SM gauge symmetry $SU(3)_c \times SU(2)_L \times U(1)_Y$  is an effective theory at the electroweak scale. It is of interest and importance to explore the existence of other gauge symmetry, in which the new force carrier(s) and particles are involved.  The extended gauge symmetries of the SM have been widely studied in the literature, such as $Z'$~\cite{Hewett:1988xc,Langacker:2008yv,Salvioni:2009mt,Chanowitz:2011ew} and $W'$~\cite{Mohapatra:1974hk,Senjanovic:1975rk,Langacker:1989xa,Rizzo:2007xs,Schmaltz:2010xr} models. Especially, if a new charged gauge boson is observed, it must be the representation of  some non-Abelian gauge group.   In this work, we thus consider  a minimal non-Abelian gauge extension of the SM and investigate the phenomenological implications at the LHC.  
 
 We extend the SM  by introducing a  new $SU(2)$ gauge symmetry, where  two new charged gauge bosons $W'^\pm$ and one neutral gauge boson $Z'$ are involved. In order to minimize the number of new particles,  a vector-like quark (VLQ) doublet of the $SU(2)$ is introduced, where new leptons are not necessary to cancel the gauge anomaly, and the $W'$ can decay into the VLQs and SM quarks;  a heavy Higgs doublet is employed  to  spontaneously break the $SU(2)$ symmetry; and a real singlet scalar is introduced to dictate the heavy Higgs production via the gluon-gluon fusion (ggF) processes. That is, there are only four new matter particles,  three new force carriers, and one new gauge coupling in this model.  We note that the new $SU(2)$ gauge symmetry is not the $SU(2)_R$, where the SM right-handed fermions belong the doublets of $SU(2)_R$. The SM particles in the model are the $SU(2)$ singlet states. 

In addition to introducing the VLQs, one can also adopt the  new representations, which are similar to  the fourth generation of the SM gauge symmetry,   to be $Q': (3,1,2)(1/6)$, $u'_R:(3,1,1)(2/3)$, $d'_R: (3,1,1)(-1/3)$, $L':(1,1,2)(-1/2)$, $e'_R:(1,1,1)(-1)$, and $\nu_R: (1,1,1)(0)$  under $(SU(3),SU(2)_1,SU(2)_2)(U(1)_Y)$. Since more new particles are involved in such model, it is expected that these new particles will lead to richer phenomena, such as more collider signatures, lepton and quark flavor physics, and $\nu_R$ can be the dark matter candidate if an unbroken $Z_2$ is imposed. Since the involving new  particles and couplings are much more than those  in the model with VLQs, in this work we focus the study on the VLQ model. 

 In order to concentrate the study on the collider signatures, we assume that only the third generation SM quarks couple to the VLQs via the Yukawa couplings. Accordingly, the charged current interactions of the SM quarks can be modified;  however,  due to the modification being  suppressed by the light quark masses, their effects can be ignored at the leading order approximation. In addition, the flavor changing neutral currents (FCNCs) happen between the third generation quarks and VLQs,  thus, the influence on the low energy flavor physics  is small.

From the electroweak $\rho$-parameter precision measurement,  it is found that the $W'$ and $Z'$ gauge bosons have to be heavier than 1 TeV. If we further assume that the VLQs are heavier than the heavy Higgs boson,  the heavy Higgs particle can only decay through the loop effects.  It is known that the loop integral for a scalar to diphoton decay strongly depends on the spin property of a particle in the loop; for instance, the ratios of loop integrals for spin-$0$, -$1/2$, and -$1$ particles are  $A_0: A_{1/2}:A_{1} \sim 1/3 : 4/3 : 7$~\cite{Gunion:1989we}. Clearly, despite  the magnitudes of  the  couplings involved,  it is more efficient to enhance the BR of  the  diphoton decay if   new spin-$1/2$ or/and -$1$ particles can make the contributions.  Hence, the  $W'$ and VLQs  in the model play an important role in the properties of the heavy Higgs boson.

The paper is organized as follows. In Sec.~II, we introduce the model.  In Sec.~III, we study the constraints on the new gauge coupling and masses of new gauge bosons, and analyze some phenomena at the LHC, such as exotic diphoton resonance and new particle production.  The summary is then  given in Sec.~IV.

\section{Model}

We start by setting  up the model. In this study, we extend the SM gauge symmetry to $SU(2)_1\times SU(2)_2\times U(1)_Y$, where the SM particles belong to the representations of $SU(2)_1 \times U(1)_Y$ and are singlets of $SU(2)_2$. To break the gauge symmetry down to $U(1)_{\rm em}$, we introduce two Higgs doublets $H_1=(2,1)_{1/2}$ and $H_2=(1,2)_{1/2}$, where the former is the SM Higgs doublet, the latter is the heavy Higgs doublet of $SU(2)_2$, and the subscripts in the representations denote the hypercharges of the Higgs doublets.  In order to minimize the number of new particles, and enhance the decays of the heavy scalar boson of $H_2$, we introduce a VLQ doublet $Q'^T=(U',D')$ of $SU(2)_2$ to the model.
In addition, we include a Higgs singlet $S'$ to produce the heavy Higgs via ggF  processes. 
Since the SM particles, Higgs doublets, and $Q'$ carry the hypercharges of $U(1)_Y$, we define the electric charges of particles to be $Q_{\rm em} = T^{(1)}_3 + T^{(2)}_3 + Y$, where $T^{(1,2)}_3 = \sigma_3/2$ and $\sigma_3$ is the diagonalized Pauli matrix. Accordingly, the electric charges of $U'$ and $D'$ are $2/3$ and $-1/3$, respectively.   For clarity, we show the representations and charge assignments of particles under the gauge symmetry of $SU(3)_C \times SU(2)_1 \times SU(2)_2 \times U(1)_Y$ in Table~\ref{tab:charge}.  

\begin{table}[hpbt]
\begin{center}
\caption{Representations and  charge assignments of particles under the gauge 
symmetry $SU(3)_C\times SU(2)_1\times SU(2)_2 \times U(1)_Y$ , where $Q'$ denotes the vector-like quark, and both left- and right-handed states carry the same charges.  }
\label{tab:charge}
\begin{ruledtabular}
\begin{tabular}{c||ccccccc||ccc}
&\multicolumn{7}{c||}{Fermions} & \multicolumn{3}{c}{Scalar} \\ \hline
& ~$Q_L$~ & ~$u_R^{}$~ & ~$d^{}_{R}$ ~ & ~$L_L$~  & ~$e_R$~ & ~$Q'_{L(R)}$  & ~$H_1$ & ~$H_2$~ & ~$S'$~ \\\hline 
$SU(3)_C$ & $\bf{3}$ & $\bf{3}$  & $\bf{3}$ & $\bf{1}$ & $\bf{1}$ & $\bf{3}$     & $\bf{1}$ & $\bf{1}$   & $\bf{1}$ \\\hline 
$SU(2)_1$ & $\bf{2}$ & $\bf{1}$  & $\bf{1}$ & $\bf{2}$ & $\bf{1}$ & $\bf{1}$     & $\bf{2}$ & $\bf{1}$   & $\bf{1}$ \\\hline 
$SU(2)_2$ & $\bf{1}$ & $\bf{1}$  & $\bf{1}$ & $\bf{1}$ & $\bf{1}$ & $\bf{2}$   & $\bf{1}$ & $\bf{2}$ & $\bf{1}$  \\\hline
$U(1)_Y$ & $1/6$ & $2/3$  & $-1/3$ & $-1/2$ & $-1$ & $1/6$  & $1/2$ & $1/2$ & $0$  
\end{tabular}
\end{ruledtabular}
\end{center}
\end{table}

Although the introduced new particles belong to the representations of $SU(2)_2$, they can couple to the SM particles through the mixings  from the  Yukawa sector, scalar potential, and gauge sector. To derive these new interactions,  we first study the Yukawa sector and scalar potential that dictates the SSB. Hence, we write them as:
 \begin{eqnarray}
- {\cal L}  &=& [ y _{F} \bar Q'_L Q'_R S' + y_b \bar Q'_L   H_2 b_R  + y_t \bar Q'_L \tilde H_2  t_R \nonumber \\
&+&   m_\Psi \bar Q'_L Q'_R + H.c.] + V(H_1, H_2, S')\,, \label{eq:Yukawa}\\
V(H_1, H_2, S') &=& \sum_{i=1,2}\left[ \mu^2_i H^\dagger_i H_i  + \lambda_i \left( H^\dagger_i H_i \right)^2 \right] + \mu^2_S S'^2 +  \lambda_S S'^4 \nonumber \\
&& + \mu_3 S'^3  + S'(\mu_{1S} H^\dagger_1 H_1 + \mu_{2S} H^\dagger_{2} H_2) + \lambda_{12} H^\dagger_1 H_1 H^\dagger_2  H_2  \nonumber \\
&& +   \lambda_{1S}  S'^2  H^\dagger_1 H_1+ \lambda_{2S} S'^2  H^\dagger_2 H_2\,. \label{eq:V}
 \end{eqnarray}
In order to focus the study on the collider signatures, we assume that only the $b$- and $t$-quark couple to the VLQs in the Yukawa sector.
 To find the stable vacuum expectation values (VEVs) of scalar fields for SSB, we express the scalar fields as:
\begin{equation}
H_{i} = \begin{pmatrix} G^+_i \\  (v_i + h_i + i G^0_i)/\sqrt{2} \end{pmatrix}, \quad S' = (v_S + S)/\sqrt{2}\,,
\end{equation}
where $G^{\pm,0}_i$ are the unphysical Nambu-Goldstone bosons and $h_{1,2}$, and $S$ are the physical scalar bosons.   By requiring  $\partial V(v_1,v_2, v_S)/\partial v_i =0$, the minimal conditions for $v_i$ are obtained as:
   \begin{eqnarray}
 &&  \mu^2_1 + \lambda_1 v^2_1 + \frac{1}{2} ( \lambda_{12} v^2_2 + \lambda_{1S} v^2_S) + \frac{\mu_{1S} }{\sqrt{2}}v_S=0\,,  \nonumber \\
 && \mu^2_2 + \lambda_2 v^2_2 + \frac{1}{2} ( \lambda_{12} v^2_1 + \lambda_{2S} v^2_S) + \frac{\mu_{2S} }{\sqrt{2}} v_S =0 \,,  \nonumber \\
 && \mu^2_S v_S+ \lambda_S v^3_S + \frac{v_S}{2} (\lambda_{1S} v^2_1 + \lambda_{2S} v^2_2 ) + \frac{3 \mu_S}{2\sqrt{2}} v^2_S  + \frac{1}{2\sqrt{2}} \left( \mu_{1S} v^2_1 + \mu_{2S} v^2_2 \right)=0 \,.
   \end{eqnarray}
The mass-square matrix for the scalar bosons, which satisfies above conditions, can thus be expressed as:
  \begin{eqnarray}
  M^2 &=& \left(\begin{array}{ccc}
                    m^2_{h_1} & \lambda_{12} v_1 v_2  & \lambda_{1S} v_1 v_S + \frac{\mu_{1S} v_1}{\sqrt{2}} \\
                    \lambda_{12} v_1 v_2 & m^2_{h_2}  & \lambda_{2S} v_2 v_S + \frac{\mu_{2S} v_2}{\sqrt{2}} \\
                    \lambda_{1S} v_1 v_S + \frac{\mu_{1S} v_1}{\sqrt{2}} & \lambda_{2S} v_2 v_S +  \frac{\mu_{2S} v_2}{\sqrt{2}} & m^2_S
                     \end{array} \right) \,,   \end{eqnarray}
  where the diagonal elements are  $m^2_{h_1} = 2 \lambda_1 v^2_1$, $m^2_{h_2}=2 \lambda_1 v^2_2$, and 
  \begin{equation}  
  m^2_S = 2 \lambda_S v^2_S+ \frac{3\mu_S v_S}{2\sqrt{2}}  - \frac{\mu_{1S} v^2_1 + \mu_{2S} v^2_2 }{2 \sqrt{2} v_S} \,. 
  \end{equation}
   It is clear that the parameters $\lambda_{12}$ and  $\lambda_{1S}(\mu_{1S})$ control the mixtures of $h_1$-$h_2$ and $h_1$-$S$, respectively. Since $S$ field directly couples to the heavy VLQs,  any sizable  mixings between $h_1$ and $(h_2, S)$ may cause too large  Higgs production cross section and BR for the Higgs to diphoton decay;
  for instance, the diphoton signal strength parameter, defined by $\mu^{\gamma\gamma}_i =[\sigma(pp\to h)/\sigma(pp\to h)_{\rm SM}] \cdot [\text{BR}(h\to \gamma\gamma)/ \text{BR}(h\to \gamma\gamma)_{\rm SM}] \equiv \mu_i \cdot \mu_f$, 
 would conflict with the data which are measured  by ATLAS~\cite{Aad:2015gba} and CMS~\cite{Khachatryan:2016vau} and  show $\mu_i^{\gamma \gamma} = 1.17 \pm 0.28$ and $1.11^{+0.25}_{-0.23}$, respectively.  
 For this phenomenological reason, we adopt $\lambda_{12}, \lambda_{1S}, \mu_{1S}/m_S \ll 1$. 
 Therefore,  in this model, $h_1$ is regarded as the SM Higgs $h$;
 \begin{equation}
 h_1 \simeq h, \quad v \equiv v_1, \quad m_{h} \equiv m_{h_1} \simeq \sqrt{2 \lambda_1} v,
 \end{equation}
where $v \simeq 246$ GeV is the VEV of SM Higgs, and we use $h$ instead of $h_1$ hereafter. 
   As a result,  we only need to focus on  a $2\times 2$ matrix, expressed as:
   \begin{eqnarray}
   M^2_{h_2 S} =  \left(\begin{array}{cc}
                     m^2_{h_2}  & m^2_{23} \\
                    m^2_{23} & m^2_S
                     \end{array} \right) 
   \end{eqnarray}
  with $m^2_{23} =\lambda_{2S} v_2 v_S + v_2 \mu_{2S} /\sqrt{2}$. Accordingly, the physical masses are given by:
   \begin{eqnarray}
   m^2_{H/H_S} &=& \frac{m^2_{S} +m^2_{h_2} }{2} \mp \frac{1}{2} \sqrt{\left( m^2_S -m^2_{h_2} \right)^2 + 4 m^4_{23} }\,.
      \end{eqnarray}
    The relationship between  physical   and weak states is parametrized as:
     \begin{eqnarray}
     \left( \begin{array}{c}
      H \\
      H_S  \end{array} \right) = \left( \begin{array}{cc}
      \cos\phi & -\sin\phi \\
      \sin\phi & \cos\phi
      \end{array} \right) \left( \begin{array}{c}
      h_2 \\
      S  \end{array} \right) \,, \label{eq:mixing}
     \end{eqnarray}
     where the mixing angle is given by $\sin\phi = \left( 1-\sqrt{1-\sin^22\phi} \right)^{1/2} $ and $\sin2\phi = 2 m^2_{23}/(m^2_{H_S}  - m^2_{H} )$. 
      $H$ and $H_S$ are the new   heavy scalar bosons. Since the $H$ dictates the $SU(2)_2$ breaking, hereafter we name it as the heavy Higgs boson.  From the scalar potential of Eq.~(\ref{eq:V}), it can be seen that twelve  parameters are introduced. Ignoring the small $\lambda_{12,1S}$, and $\mu_{1S}/m_S$, the number of relevant free parameters is eight. Since $\mu_{2S}$ appears in $m^2_{23}$ and $m^2_S$, its information cannot be extracted singly. In the current  numerical analysis, we set $\mu_{2S} =0$ for simplicity. In terms of VEVs, masses of scalar bosons, and mixing angle,  the set of free parameters from the scalar potential is chosen as: $v_{1,2}$, $v_S$, $m_{h, H}$,  $m_{H_S}$, and $\sin\phi$. If we take $v\equiv v_1 \approx 246$ GeV and  $m_h \approx 125$ GeV, the undetermined  free parameters in scalar sector are $v_S$, $m_{H, H_S}$, and $\sin\phi$.
  
  After SSB, all fermions are in physical states. Since only the third generation of  quarks couples to the VLQs and $H_2$ doublet, we can choose the basis for which  the first two generations of quarks are in mass eigenstates; however, the Dirac mass matrix for $t-U'$ and $b-D'$ quarks can be formulated by: 
  \begin{eqnarray}
         M_q &=& \left( \begin{array}{cc}
                   \bar m_q & 0 \\
                   m_{qQ} & m_F \end{array} \right)\, \label{eq:qmass}
  \end{eqnarray}
  where  $q=t (b)$ and $Q=U'(D')$ quarks, $\bar m_q$ is the mass of the SM quark $q$ before introducing  the VLQs, $m_F= m_\Psi + y_F v_S/\sqrt{2} $, and $m_{qQ}= y_q v_2/\sqrt{2}$.  We note that $m_{U'} = m_{D'} = m_F$. $M_q$ can be diagonalized by a bi-unitary transformation $M^{\rm dia}_q = V^q_L M_q V^\dagger_R$, where $V^q_{L, R}$ are unitary matrices. The $V^q_L$ and $V^q_R$ can be obtained through 
  $M^{\rm dia}_q M^{\rm dia \dagger}_q = V^q_L M_q M^\dagger_q V^{q\dagger}_L$ and $M^{\rm dia\dagger }_q M^{\rm dia }_q = V^q_R M^\dagger_q M_q V^{q\dagger}_L$, respectively.  If we parametrize the $V^q_{L(R)}$ to be a $2\times 2$ matrix, as shown in Eq.~(\ref{eq:mixing}), and use the angle $\theta^q_{L(R)}$ instead of $\phi$,  with $\bar m_q, m_{qQ} < m_F$ we find:
  \begin{eqnarray}
  \tan\theta^q_L \approx \frac{\bar m_q m_{qQ} }{m^2_F}\,, \ \ \ \tan\theta^q_R \approx \frac{m_{qQ}}{m_F}\,. \label{eq:q_ang}
  \end{eqnarray}
Note that the SM quark mass without VLQs is given by $\bar m_q = y^{\rm diag}_q v /\sqrt{2}$ where $y^{\rm diag}_q$ is the component of diagonalized SM Yukawa coupling matrix.  
In the following analysis, we use the notations of $T$ and $B$ to present  the physical states of $U'$ and $D'$, respectively.  If the new exotic quarks are as heavy as ${\cal O}$(TeV),  the masses of the quarks  can be simplified as $m_t \approx \bar m_t$, $m_b \approx \bar m_b$, $m_T \approx m_B \approx m_F$.  We use these simple relations for the numerical calculations and phenomenological analysis. The Yukawa couplings of $(S, h_2, h)$ to quarks are thus  presented in Table~\ref{tab:Yukawa}, where $s^q_{L(R)}=\sin\theta^q_{L(R)}$, $c^q_{L(R)}= \cos\theta^q_{L(R)}$, $q$ is the SM quark $t$ or $b$, and $Q$ stands for the VLQ $T$ or $B$. 
  
\begin{table}[phtb]
\begin{center}
\caption{ Yukawa couplings of scalar bosons to quarks.} 
\label{tab:Yukawa}
\begin{ruledtabular}
\begin{tabular}{ccccc} 
 Field & $\bar Q[...] Q$ &  $\bar q [...] Q$ & $\bar Q[...] q $ &  $\bar q[...] q$  \\ \hline
  $S$     & $-\frac{y_F}{\sqrt{2}} c^q_L c^q_R$ & $ \frac{y_F}{\sqrt{2}} (s^q_L c^q_R P_R + c^q_L s^q_R P_L)$ & $ \frac{y_F}{\sqrt{2}} (s^q_L c^q_R P_L + c^q_L s^q_R P_R )$ & $-\frac{y_F}{\sqrt{2}} s^q_L s^q_R$ \\ \hline
 $h_2$ & $-\frac{y_q}{\sqrt{2}} c^q_L s^q_L$ & $-\frac{y_q}{\sqrt{2}} (-s^q_L s^q_R P_R + c^q_L c^q_R P_L)$ & $-\frac{y_q}{\sqrt{2}} (-s^q_L s^q_R P_L + c^q_L c^q_R P_R)$ & $\frac{y_q}{\sqrt{2}} s^q_L c^q_R$ \\ \hline
 $h$ & $-\frac{m_q}{v}s^q_L s^q_R$ & $-\frac{m_q}{v} ( c^q_L s^q_R P_R + s^q_L c^q_R P_L)$ & $-\frac{m_q}{v} ( c^q_L s^q_R P_L + s^q_L c^q_R P_R)$ & $-\frac{m_q}{v} c^q_L c^q_R$
\end{tabular}
\end{ruledtabular}
\end{center}
\end{table}

 To get the gauge interactions in the model, we write the covariant derivative as:
\begin{equation}
D_\mu  = \left( \partial_\mu - i  g_i T^{(i)}_a  A^a_{i\mu} - i  g_Y Y B_\mu \right)\,, \label{eq:cov}
\end{equation}
where $g_i$ and $A^a_{i\mu}$ ($a=1$-$3$)  are the gauge coupling and gauge fields of SU(2)$_i$, $g_Y$ and $B_\mu$ are the gauge coupling and gauge field of $U(1)_Y$, $T^{(i)}_a = \sigma_a/2$ and $\sigma_a$ are  the Pauli matrices, and $Y$ is the hypercharge of a particle. The masses of gauge bosons and the couplings of $h$ and $h_2$ to gauge bosons are dictated by the kinetic terms of the $H_{1}$ and $H_2$ fields, which are defined by $(D_\mu H_i)^\dagger (D^\mu H_i)$. Using Eq.~(\ref{eq:cov}),  the covariant derivative of $H_i$ can be written as:
 \begin{equation}
 D_\mu H_i \supset \left( \begin{array}{cc} 
       g_i A^3_{i\mu}/2 + g_Y B_\mu/2 & g_i W^+_{i\mu}/\sqrt{2} \\
       g_i W^-_{i\mu}/\sqrt{2}  &  -g_i A^3_{i\mu}/2 + g_YB_\mu /2 \end{array} \right) 
       \left(\begin{array}{c} 
        0 \\
        (v_i + h_i)/\sqrt{2} \end{array} \right)\,, \label{eq:covH}
 \end{equation}
 where the charged gauge fields are defined by $W^\pm_i = (A^1_i \mp i A^2_i)/\sqrt{2}$. Since the gauge transformations of $H_1$ and $H_2$ are independent,  $W^\pm_1$ and $W^\pm_2$ do not mix with each other.  One can thus name them as the SM and new charged gauge bosons $W^\pm$ and $W'^\pm$, and their masses can be easily obtained as $m_W = g v/2$ and $m_{W'} = g_2 v_2/2$, respectively, where we have used $g$ and $v$ instead of $g_1$ and $v_1$. From Eq.~(\ref{eq:covH}), the triple couplings of $h_i$ and $W^\pm_{i\mu}$ can be expressed as:
 \begin{eqnarray}
 {\cal L}_{h_i WW} = g m_W h W^+_\mu W^{-\mu} + g_2 m_{W'} h_2 W'^+_\mu W'^{-\mu}\,.
 \end{eqnarray}
  
 Unlike the charged gauge bosons, both $H_1$ and $H_2$ carry $U(1)_Y$ charge. When $SU(2)_1\times SU(2)_2 \times U(1)_Y$ breaks to $U(1)_{\rm em}$, the gauge fields $A^3_{1\mu}$, $A^3_{2\mu}$, and $B_{\mu}$ of $U(1)_Y$  mix so that we have two massive neutral gauge bosons $Z$ and $Z'$ and one massless photon.   
 The mass-square matrix for the neutral gauge boson is  expressed as: 
 \begin{equation}
 {\cal L}_M = \frac{1}{8} \left( \begin{array}{c} A_{2 \mu}^3 \\ A_{1 \mu}^3 \\ B_\mu \end{array} \right)^T
\left( \begin{array}{ccc} v_2^2 g_2^2 & 0 & - v_2^2 g_2 g_Y \\ 0 & v_1^2 g^2 & - v_1^2 g g_Y \\ - v_2^2 g_2 g_Y & -v_1^2 g g_Y & (v_1^2 + v_2^2) g_Y^2 \end{array} \right)
\left(  \begin{array}{c} A_{2}^{3 \mu} \\ A_{1}^{3 \mu} \\ B^\mu \end{array} \right). \label{eq:m2_ZA}
  \end{equation}
Since $U(1)_{\rm em}$ symmetry is preserved, to show the massless photon state, we adopt the basis of gauge fields as: 
\begin{equation}
\left(  \begin{array}{c} A_{2 \mu}^{3} \\ A_{1 \mu}^{3} \\ B_\mu \end{array} \right) =
\left( \begin{array}{ccc} c_\theta & 0 &  s_\theta \\ 0 & 1 & 0 \\ -s_\theta & 0 & c_\theta \end{array} \right)
\left(  \begin{array}{ccc} 1 & 0 & 0 \\ 0 & c_W & s_W \\ 0 & -s_W & c_W \end{array} \right) 
 \left( \begin{array}{c} Z_{2 \mu} \\ Z_{1 \mu} \\ A_\mu \end{array} \right)  \,,
  \label{eq:transform}
\end{equation} 
  where $s_\theta =\sin\theta = g_Y /\sqrt{g^2_2 + g^2_Y}$, $c_\theta=\cos\theta = g_2/\sqrt{g^2_2 + g^2_Y}$, $g'=g_Y c_\theta$, $s_W=\sin\theta_W=g'/\sqrt{g^2 + g'^2}$, $c_W = \cos\theta_W = g/\sqrt{g^2 + g'^2}$, $\theta_W$ is the Weinberg's angle in the SM, and $A_\mu$ is the massless photon. In terms of this basis, the mass-square matrix of Eq.~(\ref{eq:m2_ZA}) is reduced  to be a $2 \times 2$  matrix, which  is just  for  the $Z_1$ and $Z_2$ gauge bosons. 
  Since  the gauge coupling $g_2$  is the only new free parameter in the gauge sector, the  $s_\theta$, $c_\theta$, and $g_Y$ can be expressed by the gauge couplings $g_2$ and $g'$ as:
  \begin{eqnarray}
  c_\theta &=& \sqrt{1 - g'^2/g^2_2}\,, \  \  s_\theta = g'/g_2\,,  \nonumber \\
  g_Y &=& \frac{g_2 g'}{\sqrt{g^2_2 - g'^2}}\  \  \ \text{with } g' < g_2\,. 
  \end{eqnarray}
Under the basis in Eq.~(\ref{eq:transform}), 
the mass-square matrix for the massive gauge bosons $Z_1$ and $Z_2$ is given by:
  \begin{eqnarray}
  M^2_{Z_1 Z_2} &= &\left( \begin{array}{cc}
                   m^2_{Z_1}  & m^2_{Z_1Z_2} \\
                   m^2_{Z_1Z_2} & m^2_{Z_2} 
     \end{array} \right) \,,  \label{eq:m2Z} \\
  m^2_{Z_2} &=& \frac{v^2_2 g^4_2 + v^2 g'^4 }{4 (g^2_2 - g'^2)}  \,,  \ \  m^2_{Z_1} = \frac{v^2}{4} (g^2 + g'^2)\,,\nonumber \\
 m^2_{Z_1 Z_2} &=& \frac{v^2 g'^2}{4} \sqrt{\frac{g^2 + g'^2}{g^2_2 -g'^2}}\,.  \nonumber 
  \end{eqnarray}
 As a result, the masses of $Z$ and $Z'$ and their mixing angle can be written as:
  \begin{eqnarray}
  m^2_{Z/Z'} &=& \frac{m^2_{Z_1} + m^2_{Z_2}}{2} \pm \frac{1}{2} \sqrt{(m^2_{Z_2} - m^2_{Z_1})^2 + 4 m^4_{Z_1 Z_2}} \,, \\ \nonumber
  \sin2\theta_Z &=& \frac{2m^2_{Z_1 Z_2}}{ m^2_{Z'} - m^2_{Z}}\,, 
  \end{eqnarray}

 It is known that the $\rho$-parameter in the SM is $\rho = m_W/(m_Z \cos\theta_W) =1$ at the tree level, whereas the precision measurement is $\rho^{\rm exp} = 1.00040^{+0.0003}_{-0.0004}$ \cite{PDG}.  From Eq.~(\ref{eq:m2Z}),  $\rho = m^2_{Z_1}/m^2_Z$ in this model. Thus,  any sizable $m^2_{Z_1 Z_2}$ will spoil $\rho=1$. To fit the experimental bound, we have to require $m^2_{Z_1}, m^2_{Z_1 Z_2} \ll m^2_{Z_2}$.  Taking the allowed range of $\rho$ within $1\sigma$ errors, it is found that the condition to satisfy  the bound of $\rho^{\rm exp}$  is: 
  \begin{equation}
m_{Z'}  >   \sqrt{\frac{1}{4} + \frac{g'^4}{g^2_2 -g'^2}} \frac{m_Z}{\sqrt{7 \times 10^{-4}}} \,. \label{eq:limit}
 \end{equation}
Roughly, the mass of gauge boson $Z'$ has to be heavier than 1.7 TeV and the mixing angle $\theta_Z$ is of the order of $10^{-3}$. That is, the $Z$ and $Z'$ mixing effect is small and can be neglected.  Taking this approximation, the couplings of scalars to $Z$ and $Z'$ can be expressed as:
 \begin{eqnarray}
{\cal L}_{h_i ZZ} &=& \frac{1}{2} \frac{g m_{Z}}{c_W} h Z_{\mu} Z^\mu + g' t_\theta m_{Z} h Z_{\mu} Z'^\mu\nonumber \\
&+& \frac{1}{2} \left[ \frac{g^2_2}{2c^2_\theta} v_2 h_2 + \frac{g'^2 t^2_\theta}{2} v_1 h\right] Z'_{\mu} Z'^\mu
 \end{eqnarray}
with $t_\theta=\tan\theta$. 

Next, we discuss the interactions of gauge bosons and fermions. As mentioned earlier, since the symmetry breaking is dictated by the two Higgs doublets, the charged gauge bosons in $SU(2)_1$ do not mix with those in $SU(2)_2$. However, the SM quarks of $SU(2)_1$ and the VLQs of $SU(2)_2$  can  couple to $W^{'\pm}$ and  $W^\pm$ respectively through the flavor mixings, which arise from the Yukawa couplings and are shown in Eqs.~(\ref{eq:qmass}) and (\ref{eq:q_ang}). Since only the third-generation of the SM quarks mixes with the VLQs, we present the  relevant couplings of $W$ boson  to the quarks as:
 \begin{eqnarray}
 {\cal L}_{W} &=& -\frac{g}{\sqrt{2}} V_{tq'} (c^t_L \bar t_L + s^t_L  \bar T_L ) \gamma^\mu q'_L W^+_{\mu}   - \frac{g}{\sqrt{2}} V_{q'' b} \bar q''_L \gamma^\mu (c^b_L b_L + s^b_L B_L )W^+_{\mu} \nonumber \\
 &-&\frac{g_2}{\sqrt{2}}  \left(\begin{array}{c} \bar t, \bar T \end{array} \right)_L \gamma^\mu
    \left(  \begin{array}{cc}
                c^t_L c^b_L  & c^t_L s^b_L \\
                s^t_L c^b_L &  s^t_L s^b_L
   \end{array} \right) \left( \begin{array}{c} 
   b \\
   B \end{array} \right)_L W^{+}_{\mu} + H. c.\,,
 \end{eqnarray}
where $V_{qq'}$ is the Cabibbo-Kobayashi-Maskawa (CKM) matrix element, $q'=d,s$,  and $q''=u,c$.  Since both  left-handed and right-handed VLQs can couple to the $W'$-gauge boson, with the mixing angles of $\theta^q_L$ and $\theta^q_R$, the interactions of $W'$ and quarks can be formulated by:
 \begin{eqnarray}
 {\cal L}_{W_2} &=& -\frac{g_2}{\sqrt{2}}  \left(\begin{array}{c} \bar t, \bar T \end{array} \right)_\chi \gamma^\mu
    \left(  \begin{array}{cc}
                s^t_\chi s^b_\chi  & -s^t_\chi c^b_\chi \\
                -c^t_\chi s^b_\chi &  c^t_\chi c^b_\chi
   \end{array} \right) \left( \begin{array}{c} 
   b \\
   B \end{array} \right)_\chi W^{'+}_{\mu} + H. c.\,,
 \end{eqnarray}
 where $\chi$ denotes the chirality of quarks. Because we do not introduce exotic leptons in this model, the couplings of $W$-boson to the SM leptons are not changed.

 It has been shown that the neutral gauge bosons $A^3_{1\mu, 2\mu}$, and $B_\mu$ mix together when the local gauge symmetry is broken. Therefore, even without the flavor mixings of Eq.~(\ref{eq:qmass}), $Z_{1\mu}$ and $Z_{2\mu}$ can couple to VLQs and the SM quarks simultaneously.  Combining the flavor mixings $\theta^q_{L,R}$ and the gauge mixing $\theta$,  the interactions of $Z_1$ and quarks are presented as:
 \begin{eqnarray}
 {\cal L}_{Z_1} &=& -\frac{g}{c_W} \left(\begin{array}{c} \bar q, \bar Q \end{array} \right)_L \gamma^\mu
    \left(  \begin{array}{cc}
                (c^q_L)^2 T^{(1)}_3 -s^2_W Q_{\rm em} & c^q_L s^q_L T^{(1)}_3 \\
                c^q_L s^q_L T^{(1)}_3 &  (s^q_L)^2 T^{(1)}_3 -s^2_W Q_{\rm em} 
   \end{array} \right) \left( \begin{array}{c} 
   q \\
   Q \end{array} \right)_L Z_{1\mu} \nonumber \\
   &-&  \frac{g}{c_W} \left(\begin{array}{c} \bar q, \bar Q \end{array} \right)_R \gamma^\mu
    \left(  \begin{array}{cc}
                -s^2_W Q_{\rm em} & 0 \\
               0 &   -s^2_W Q_{\rm em} 
   \end{array} \right) \left( \begin{array}{c} 
   q \\
   Q \end{array} \right)_R Z_{1\mu}\,,
 \end{eqnarray}
where $q=t(b)$, $Q=T(B)$, $Q_{\rm em}$ denotes the electric charge of quark. It can be seen that the FCNCs are induced in the left-handed current interactions while the right-handed currents only have  flavor-conserving couplings. Since the mixing between $Z_1$ and $Z_2$ is small, the $Z_1$  can be regarded as the physical $Z$-gauge boson when the mixing is neglected. Similarly, one can get the $Z_2$ couplings to quarks  as follows:
 \begin{eqnarray}
 {\cal L}_{Z_2} &=& -\frac{g_2 }{c_\theta} \left(\begin{array}{c} \bar q, \bar Q \end{array} \right)_L \gamma^\mu
   {\cal M}_{2L} \left( \begin{array}{c} 
   q \\
   Q \end{array} \right)_L Z_{2 \mu} \nonumber \\
   &-&  \frac{g_2}{c_\theta} \left(\begin{array}{c} \bar q, \bar Q \end{array} \right)_R \gamma^\mu
    \left(  \begin{array}{cc}
                (s^q_R)^2 T^{(2)}_3-s^2_\theta Q_{\rm em} & -c^q_R s^q_R T^{(2)}_3 \\
             -c^q_R s^q_R T^{(2)}_3  &   (c^q_R)^2 T^{(2)}_3-s^2_\theta Q_{\rm em} 
   \end{array} \right) \left( \begin{array}{c} 
   q \\
   Q \end{array} \right)_R Z_{2\mu} \label{eq:LZ2}
 \end{eqnarray}
 with 
 \begin{equation}
{\cal M}_{2L} =  \left(  \begin{array}{cc}
                (c^q_L)^2 s^2_\theta T^{(1)}_3 + (s^q_L)^2 T^{(2)}_3-s^2_\theta Q_{\rm em} & c^q_L s^q_L (s^2_\theta T^{(1)}_3 - T^{(2)}_3 ) \\
               c^q_L s^q_L (s^2_\theta T^{(1)}_3 - T^{(2)}_3 )  &  (s^q_L)^2 s^2_\theta T^{(1)}_3 + (c^q_L)^2 T^{(2)}_3 - s^2_\theta Q_{\rm em} 
   \end{array} \right)\,. \label{eq:M2L}
 \end{equation}
These  complicated couplings can be  simplified if we adopt the limit $\theta^q_L \to 0$, which is from the result of $\theta^q_L \ll 1$.  The couplings of $Z_1$ to the SM leptons are the same as those in the SM, and thus we do not show them again. The couplings of $Z_2$ to the SM leptons are new and they are given as: 
 \begin{eqnarray}
 {\cal L}_{Z_2 \ell \ell} & = & - \frac{g_2 s^2_\theta }{2c_\theta} \left[ \bar \nu \gamma^\mu \left( \frac{1}{2}- \frac{1}{2}\gamma_5 \right)\nu  + \bar\ell \gamma^\mu \left( \frac{3}{2} + \frac{1}{2} \gamma_5 \right) \ell \right]Z_{2\mu}\,. 
 \end{eqnarray}
 In order to calculate the BRs for $h_2 \to \gamma\gamma, Z\gamma, ZZ$ decays through the $W'$-loop, and the  vertices involved are derived as:
 \begin{eqnarray}
 {\cal L}_{W'W'V} &=& \left[g_{\alpha \beta} (p^-_\mu - p^+_\mu) + g_{\mu \alpha} (p_\beta - p^-_\beta) + g_{\beta\mu} (p^+_\alpha -p_\alpha) \right] \nonumber \\
 &\times & W'^{-\alpha} W'^{+\beta} \left( g_2 c_\theta Z'^\mu_2 + e A^\mu -e\, t_W Z^\mu_1 \right)\,,
 \end{eqnarray}
 where $p^+_\alpha$, $p^-_\alpha$, and $p_\alpha$ are the momenta of $W'^+$, $W'^-$, and neutral gauge boson $Z_2/ Z_1/A$, respectively, and $t_W = \tan\theta_W$.  
 
 From Eq.~(\ref{eq:q_ang}) and with $\tan\theta^q_R < 0.3$, it can be seen that $\tan\theta^t_L < 0.05$ and $\tan\theta^b_L < 1.5 \times 10^{-3}$. It is a good approximation to ignore the contributions from $\theta^q_L$ when we focus on the leading effects. We therefore  adopt $s^q_L \approx 0$ and $c^q_L \approx 1$ in our numerical calculations. According to the result of Eq.~(\ref{eq:limit}), 
 $m_{Z'}$ has to be larger than $1.7$ GeV; unless explicitly mentioned, we fix $m_{Z'} \approx 1.8$ GeV. 
  Since we have not seen the signals of VLQ and $H_S$, in  numerical analysis we assume $m_{F,H_S} > m_H$. The other fixed values of the parameters used in the current work are summarized in Table~\ref{tab:values}. 
 
\begin{table}[phtb]
\caption{ Fixed  values of the parameters. } 
\label{tab:values}
\begin{ruledtabular}
\begin{tabular}{ccccccc} 
$v$ [GeV] & $m_h$ [GeV] &  $m_H$ [GeV] & $m_{H_S}$ [GeV] & $g$ &  $g'$ & $s^2_W$ \\ \hline
 246 & 125 & 750 & 1000 &  0.654 & 0.407 & 0.231
\end{tabular}
\end{ruledtabular}
\end{table}

\section{Phenomenology of the model}

 In this section, we discuss the constraints of the new gauge coupling and some phenomena,  such as the heavy Higgs $H$ to diphoton decay and the signature of the new particles in the model at 13 TeV LHC.

\subsection{Constraints on the new gauge coupling}

$g_2$ and $v_2$ are the two important parameters for the $H$ diphoton decay, and thus we need to study their constraints. Since the $Z'$-gauge boson can couple to the SM fermions and its mass is determined by $g_2$ and $v_2$,   it is of interest to understand  the constraints of $g_2$ and $m_{Z'}$ from the dijet and dilepton experiments at the LHC.  It is found that the constraints from dijet channels are not as strong as those from dileptons, and thus we  focus on the dilepton channels. 
 
\begin{figure}[t] 
\includegraphics[width=70mm]{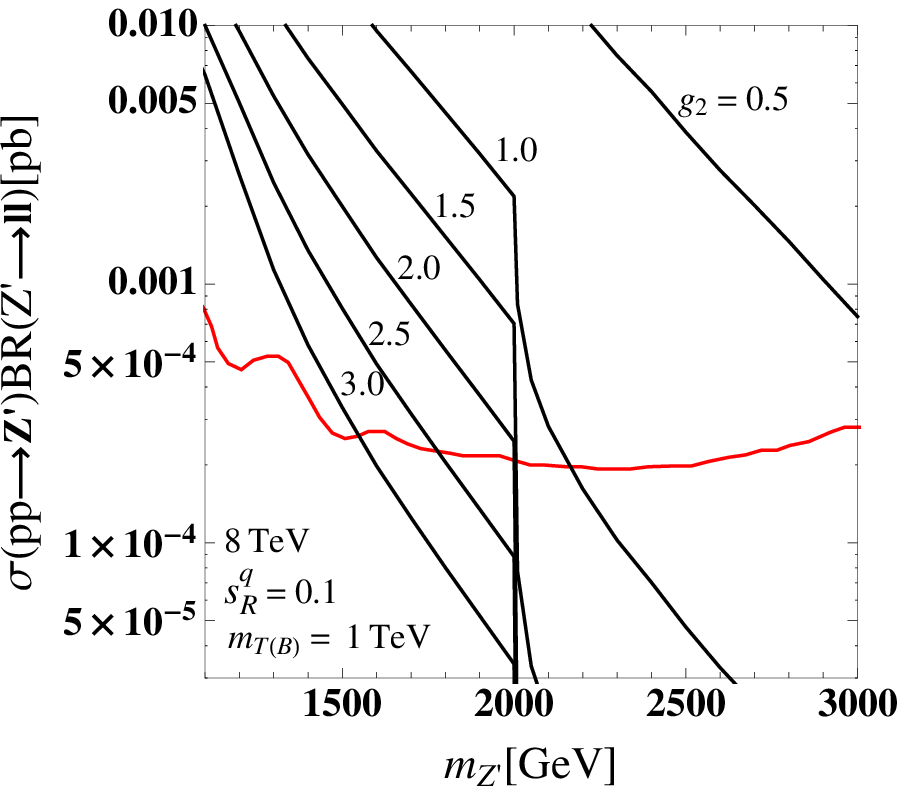} 
\includegraphics[width=70mm]{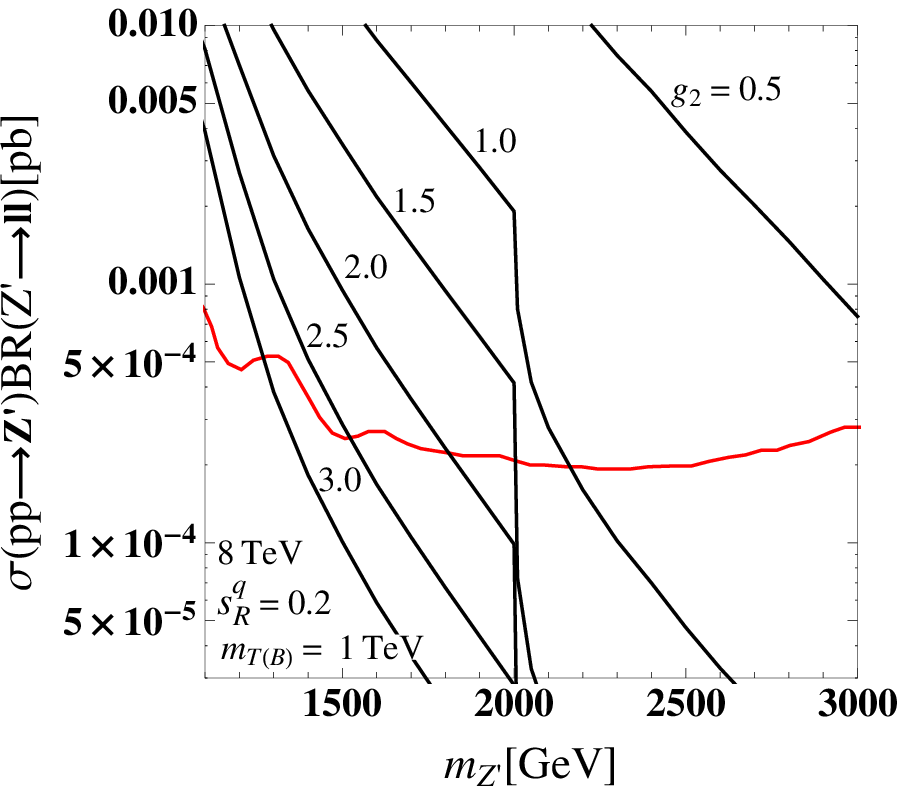} 
%
\caption{  $\sigma(pp \to Z') Br(Z' \to \ell^+ \ell^-)$ as a function of $m_{Z'}$ at $\sqrt{s}=8$ TeV with various values of $g_2$ for $s^q_R=0.1$ (left) and $s^q_R=0.2$ (right), where the masses of VLQs are fixed to be 1 TeV and the dashed red line shows the upper limit from the ATLAS experiment~\cite{Aad:2014cka}. 
\label{fig:CXBR8}}
\end{figure}

\begin{figure}[t] 
\includegraphics[width=70mm]{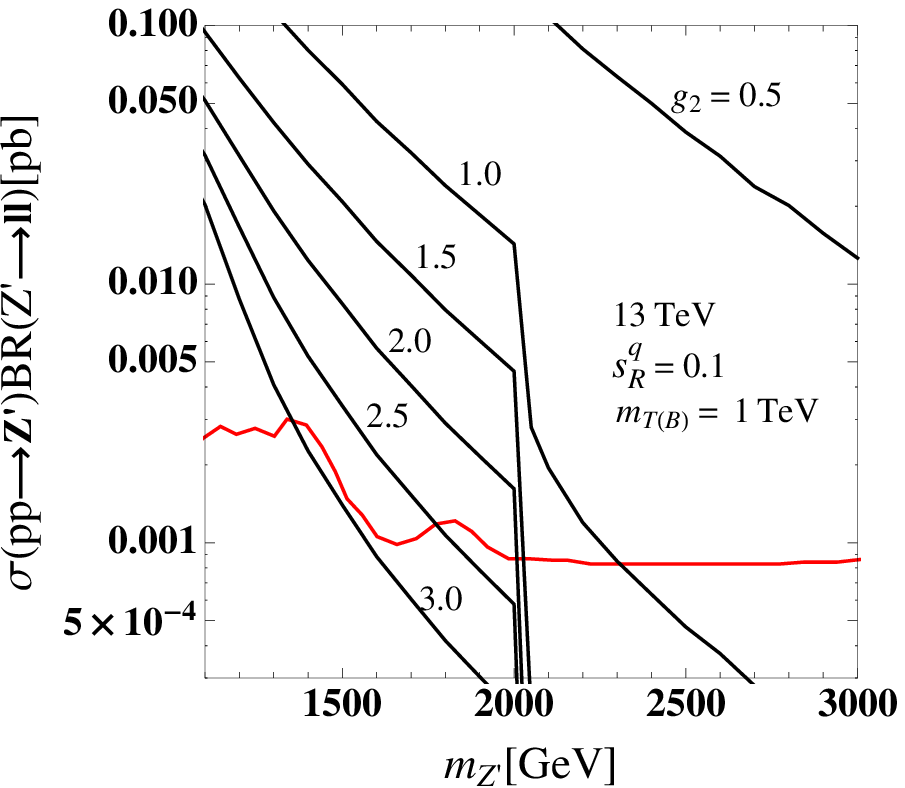} 
\includegraphics[width=70mm]{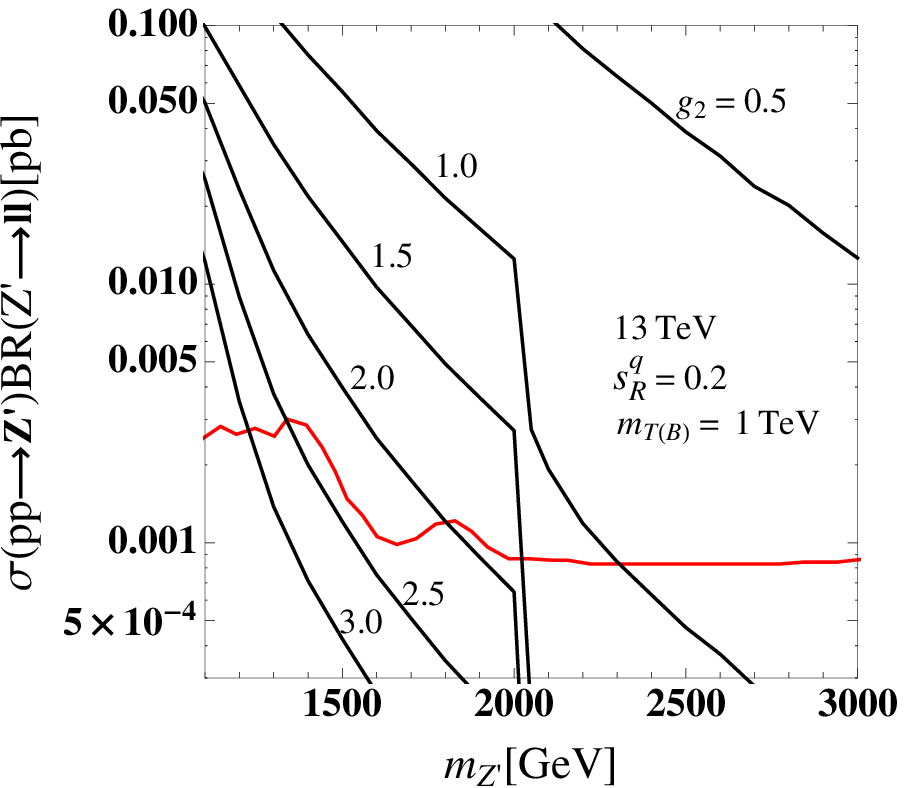} 
%
\caption{  The legend is the same as that in Fig.~\ref{fig:CXBR8} but for $\sqrt{s}=13$ TeV, where the experimental bound is from the ATLAS measurements~\cite{ATLAS:dilepton}.} \label{fig:CXBR13}
\end{figure}

 In order to  calculate the production cross section for $pp\to Z'\to \ell^+ \ell^-$ ($\ell=e,\mu$), we implement the vertices of our model into  CalcHEP~\cite{Belyaev:2012qa} and use the CTEQ6L ~\cite{Nadolsky:2008zw} parton distribution functions (PDFs).  With the interactions derived earlier, the production cross section for $pp\to Z'\to \ell^+ \ell^-$ as a function of $m_{Z'}$ is presented  in Fig.~\ref{fig:CXBR8}, where the left (right) panel is for $s^q_R=0.1 (0.2)$ at $\sqrt{s}=8$ TeV, the different lines denote the different values of $g_2$, and the masses of VLQs have been fixed to be $m_F=1$ TeV. The dashed red lines in the plots are the  bound from the  ATLAS experiments~\cite{Aad:2014cka}.  It can be seen that the cross section is decreasing when $g_2$ is increasing, and this  can be  ascribed to the couplings $Z'$-$q$-$q$ that depend on $s_\theta^2 g_2= g'^2/g_2$. The discontinuity at $m_{Z'} = 2$ TeV shows that the decay channels  of $Z' \to \bar T T, \bar B B$ are open. Besides the results at $\sqrt{s}=8$ TeV, we also show the results at $\sqrt{s}=13$ TeV in Fig.~\ref{fig:CXBR13}, where the experimental bound is from the ATLAS measurements~\cite{ATLAS:dilepton}. It is clear that the 13 TeV data have a slightly stronger constraint than the 8 TeV data. From the plots, it can be seen that a larger $s^q_R$ can weaken the constraint because  the BRs for $Z' \to T t, B b$ are enhanced; that is, the BR for $Z'\to \ell^+ \ell^-$ is relatively  suppressed. In addition, we also find  that the constraint from the $\rho$-parameter becomes dominant when $g_2 \gtrsim 2$.

\subsection{ Diphoton heavy Higgs boson decay}


With the allowed $g_2$ and $m_{Z'}$, we now study the phenomenon of the $H$ decay to diphoton. Since $H$ is a colorless scalar, the production process is through ggF.  Therefore, the effective interaction for $Hgg$ induced from the VLQ loops  is formulated as:
  \begin{eqnarray}
{\cal L}_{ggH} = -\frac{\alpha_s}{8\pi } \left( \frac{n_{F} \sin\phi c_R^q y_F }{2\sqrt{2} m_{F}} A_{1/2}(\tau)  \right) H G^{a\mu \nu}G^a_{\mu \nu} \,,\label{eq:LggS}
 \end{eqnarray}
where $n_{F}= 2$ is the number of  VLQs  and the loop function is:
 \begin{equation}
A_{1/2}(\tau) =  -2 \tau [1+(1-\tau) f(\tau)^2]
  \end{equation}
  with $\tau= 4 m_{F}^2/m_H^2$ and $f(x)=\sin^{-1}(1/\sqrt{x})$.  Using Eq.~(\ref{eq:LggS}), we can directly calculate the $H$ production cross section. Since we take $m_{F, H_S} > m_H$ and $\lambda_{12}, \lambda_{1S}, \mu_{1S}/m_S \ll 1$,  the main  $H$ decays are $H\to gg,\gamma\gamma, Z\gamma$. Although $H$ decay to $t$- and $b$-quark is allowed, due to the suppression of flavor mixings and $m_{t(b)}/m_F$, the associated BRs are much smaller than that of the diphoton decay. As such, we concentrate on the decays $H\to gg, \gamma\gamma, Z\gamma$ in the calculations. 
  
 From Eq.~(\ref{eq:LggS}), the partial decay width for $H \to gg$ is derived as:
  \begin{equation}
\Gamma(H \rightarrow gg)  =\frac{ \alpha_s^2  m_H^3 }{32 \pi^3}  \left| \frac{n_{F} \sin\phi c^q_R y_F}{2 \sqrt{2} m_{F}}    A_{1/2}(\tau) \right|^2\,. \label{eq:Gamma_gg}
\end{equation}
It can be seen that $\Gamma(H\to gg)$ strongly depends on the  $y_F$, $\sin\phi$, $m_F$, and flavor mixing $s^q_R$. In addition to  the VLQ loops, the $W'$-loop also contributes to $H \to \gamma\gamma$. The partial decay width for $H\to \gamma\gamma$ can be expressed as: 
  \begin{equation}
  \Gamma(H \to \gamma\gamma) = \frac{\alpha^2 m^3_H}{256 \pi^3}  \left| - \frac{y_F \sin \phi c_R^q Q^2_{{\rm em}} N_c}{\sqrt{2} m_{F}} A_{1/2}(\tau) + \frac{g_2 \cos \phi}{2 m_{W'}} A_{1}(\xi) \right|^2\,, \label{eq:Hgaga}
  \end{equation} 
where $N_c=3$ is the number of colors, $Q^2_{{\rm em}}=5/9$ is the sum of electric charge squares of $T$ and $B$ quarks, and $\xi = 4 m_{W'}^2/m_H^2$.
The loop function for the $W'$-gauge boson  is given by:
\begin{equation}
A_{1}(\xi) = 2+ 3 \xi + 3\xi (2- \xi) f(\xi)^2.
\end{equation}
Since the $W'$ contribution in Eq.~(\ref{eq:Hgaga}) is suppressed by $m_W'$, $H\to gg$ is the dominant decay mode. Due to $\Gamma(H\to gg) \gg \Gamma(H\to \gamma\gamma)$ and $\Gamma(H\to Z\gamma) \approx  \Gamma(H\to \gamma\gamma)$, we do not show up the detailed formula for $\Gamma(H\to Z\gamma)$, however,  we include its numerical value when calculating the width. Although $H\to ZZ$ decay is allowed in our model,  due to $\Gamma(H\to ZZ) \ll \Gamma(H\to \gamma\gamma)$, we ignore its contribution. 

 Since the $H$ production is dominated by the ggF channel, the diphoton  production cross section at the center-of-energy of $\sqrt{s}$ in the narrow width approximation can be expressed as~\cite{Franceschini:2015kwy}:
 \begin{eqnarray}
 \sigma(gg \to H \to \gamma\gamma) \approx  \frac{C_{gg}}{s} \frac{\Gamma_{gg}}{m_H} {\cal B}_{\gamma\gamma}\,, \label{eq:ggHgaga}
 \end{eqnarray}
where  $C_{gg}$ is related to the gluon luminosity function, $\Gamma_{gg} \equiv \Gamma(H\to gg)$, and ${\cal B}_{\gamma\gamma}\equiv BR(H \to \gamma\gamma)$ is the BR for the decay $H \to \gamma\gamma$. 
In order to perform the numerical analysis,   we adopt $C_{gg}\approx 2137$ at $\sqrt{s}=13$ TeV and  the K-factor for gluon fusion production process as $K_{gg} \simeq 1.5$~\cite{Franceschini:2015kwy}.  For comparison with the current upper bound on the diphoton resonance,  we take the ATLAS data with $1\sigma$ errors and $m_H=750$ GeV~\cite{ATLAS:2016eeo} as:
\begin{equation}
\label{eq:CXdiphoton}
\sigma(gg \to H \to \gamma \gamma) \leq 1.2 \, {\rm fb}\,.
\end{equation}

\begin{figure}[tb] 
\begin{center}
\includegraphics[width=70mm]{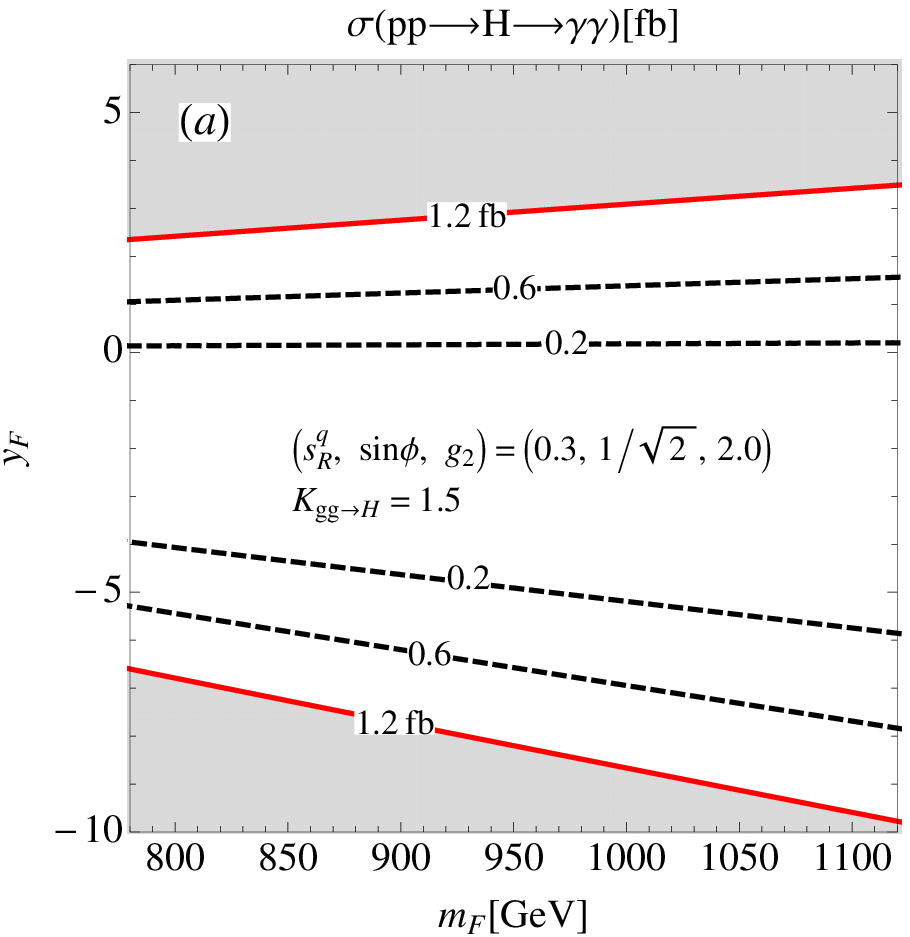}
 \includegraphics[width=70mm]{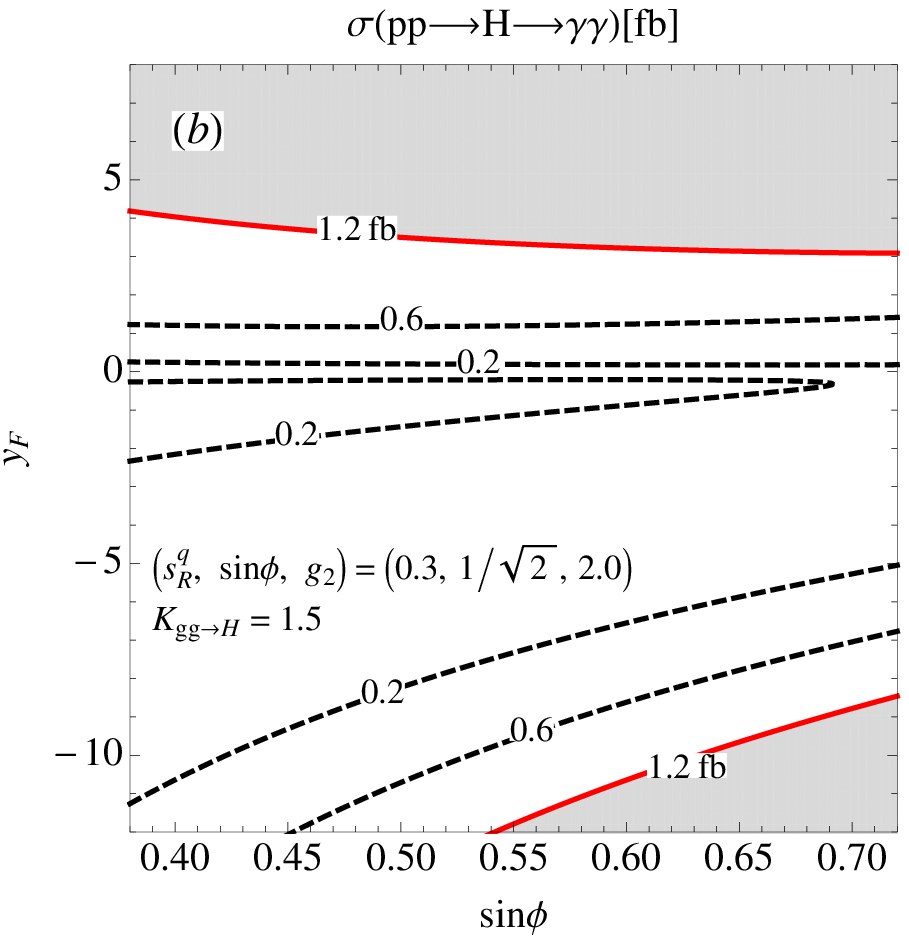}
\caption{ Contours for $\sigma(pp\to H\to \gamma\gamma)$ (dashed) as a function of (a) $y_F$ and $m_F$  (b) $y_F$ and $\sin\phi$, where  the taken values of other parameters are shown in the plots.   The gray  region is excluded by the current ATLAS data. }
\label{fig:C1}
\end{center}
\end{figure}

We now present the numerical analysis for $pp\to H\to \gamma\gamma$ by choosing some benchmark values of the free parameters. As mentioned earlier, the $H$ production and decays are sensitive to  $y_F$, $m_F$, and $\sin\phi$. In order to display the dependence of these parameters, we present the contours for $\sigma(pp \to H \to \gamma \gamma)$ at $\sqrt{s}=13$ TeV as a function of  $y_F$ and $m_F$ in Fig.~\ref{fig:C1}(a), where the dashed lines with numbers on them  are the cross section in units of fb, and we set $s^q_R=0.3$, $\sin\phi=1/\sqrt{2}$, and $g_2=3.3$. The parameter space in gray region has been excluded  by the current ATLAS data as shown in Eq.~(\ref{eq:CXdiphoton}). In addition, Fig.~\ref{fig:C1}(b) shows the contours for the cross section as a function of $y_F$ and $\sin \phi$, where  $s^q_R =0.3$, $m_F = 1$ TeV, and $g_2 =3.3$ are used.
 From the results, it can be seen that  the Yukawa coupling with $y_F > 1$ is limited by the current data. Other parameter region can be tested when more data are accumulated at the LHC.

\subsection{Collider signatures of the model}

We now study the possible collider signatures implied in the model.  It is known that the masses of $W'$ and $Z'$ have to be heavier than 1.7 TeV. The production of $W'/Z'$ pairs  is highly suppressed. For VLQ-pair production, it is found that  the $\sigma(pp\to \bar B B/ \bar T T)$ at $\sqrt{s}=13$ TeV is  around 10-80 fb when $800 \leq  m_F \leq 1100$ GeV. With $m_H < m_F < m_{W'/Z'}$, the BRs for VLQ decays are $BR(T\to  t H(h))=0.86(0.14)$ and $BR(B\to b H)\approx 1$. If we require that one $H$ decays to diphoton and the other decays to  gluon-jet,  by using the results in Fig.~\ref{fig:C1},  the cross section for $pp\to \bar Q' Q' \to \bar q q (\gamma\gamma)_H (gg)_H$ is around 0.05-0.4 fb, where $Q'=T(B)$ and $q=t(b)$. 

\begin{figure}[t] 
\includegraphics[width=50mm]{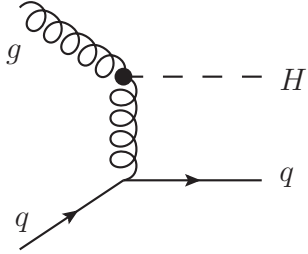}
\caption{Feynman diagram for $pp\to H+$jet process  where black dot indicates effective coupling in Eq.~(\ref{eq:LggS}). }
\label{fig:feyn_H_jet}
\end{figure}

\begin{figure}[t] 
\begin{center}
\includegraphics[width=70mm]{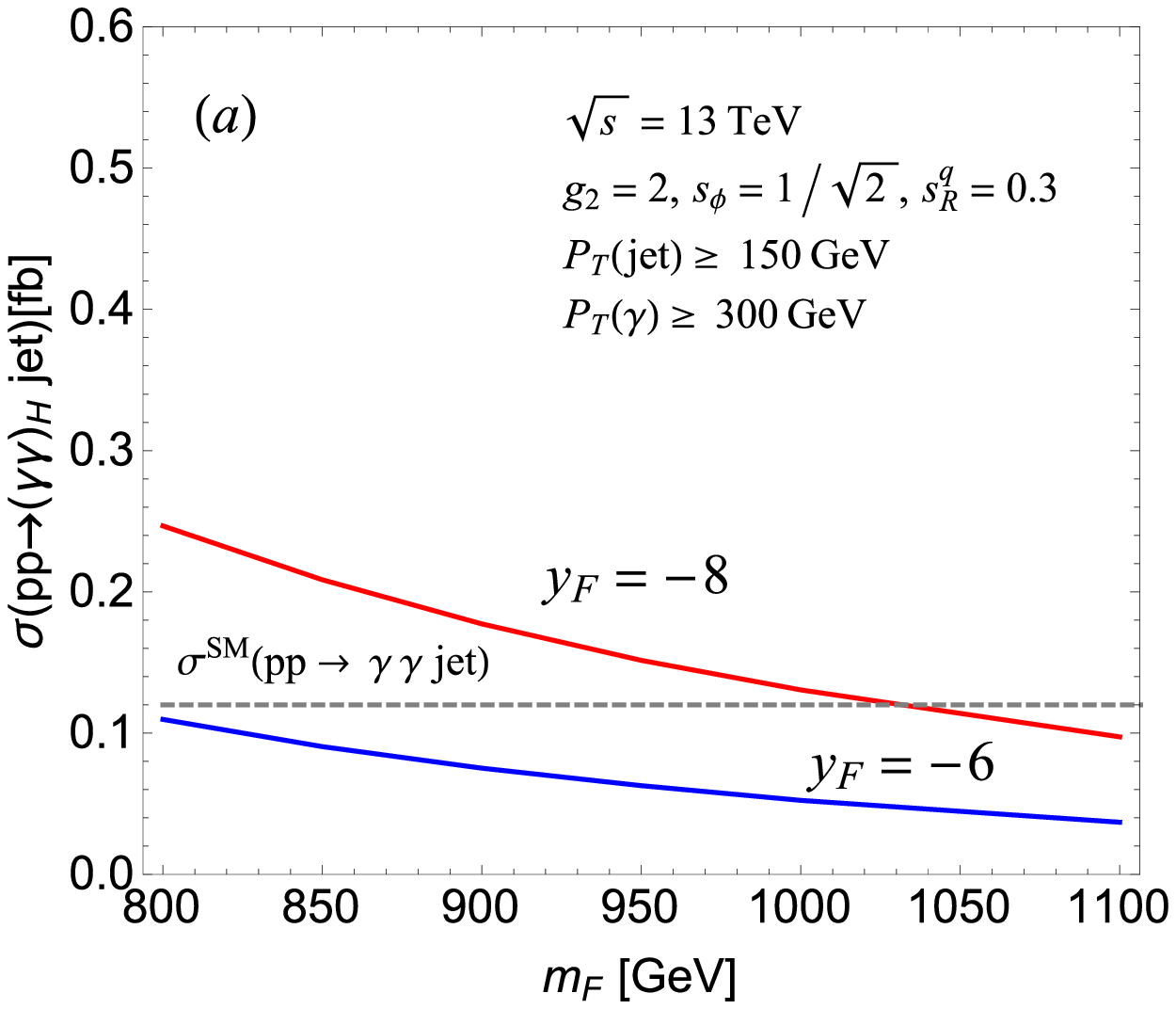}
 \includegraphics[width=70mm]{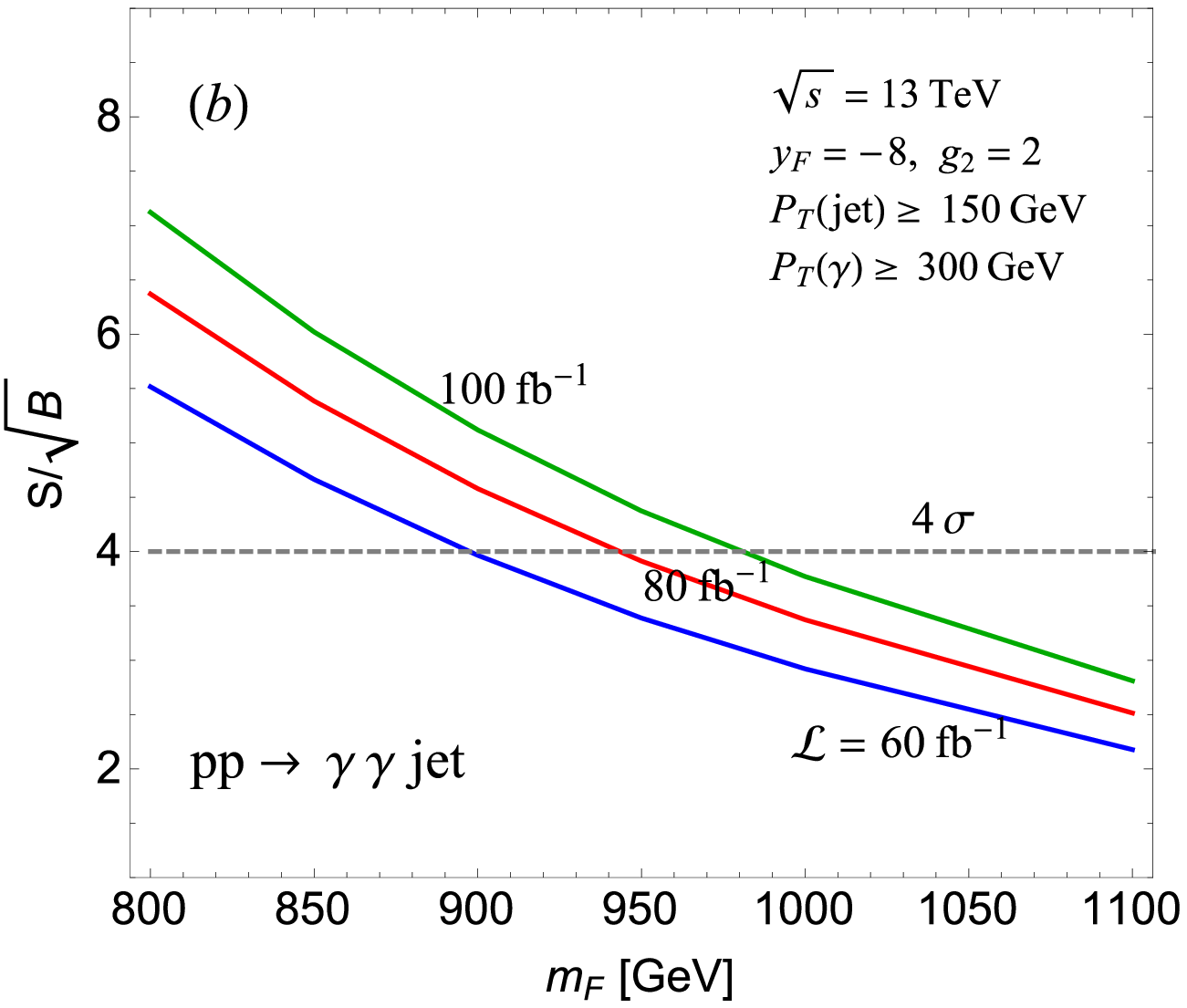}
\caption{(a) Production cross section for $pp\to (\gamma\gamma)_H$ jet as a function of $m_F$, where the dashed line is the result from the SM, the solid lines are for $y_F=-8,\,-6$, and we adopt the kinematic cuts $P_T(\rm jet)\geq 150$ GeV and $P_T(\gamma)\geq 300$ GeV to suppress the background. (b) The corresponding significance of plot (a) with $y_F=-8$ and different luminosities.  }
\label{fig:Hjet}
\end{center}
\end{figure}

Next, we study the single production of a new particle. Since $W'^\pm$  couple to the SM fermions via the flavor mixings,  the  production cross section for $W' (t,b)$  is of order of $10^{-2}$ fb. The $Z'$-boson can couple to the SM fermions without flavor mixing, however, the  production cross section for $Z' (t, b, \rm jet)$ is not large and is of order of $3$ fb. Although the VLQs can be as light as a few hundred GeV,   the single $T(B)$-quark  production  is highly suppressed. We find that the production cross section for the process $pp\to H+ {\rm jet}$ can reach 0.2 pb, where the dominant interaction is from $ggH$ shown in Eq.~(\ref{eq:LggS}) and the associated Feynman diagram is sketched in Fig.~\ref{fig:feyn_H_jet}. 
This channel can be used to further probe the scalar resonance. To show the detection possibility, we calculate the production cross section for $pp\to (\gamma\gamma)_H+ {\rm jet}$ as a function of $m_F$ in Fig.~\ref{fig:Hjet}(a), where the solid lines are for $y_F=-8,-6$, the dashed line is the SM result, the center of energy is $\sqrt{s}=13$ TeV,  and the values of the parameters are set to be $\sin\phi=1/\sqrt{2}$, $g_2=2$, and $s^q_R=0.3$. In order to suppress the contributions from the SM, we adopt the following kinematic cuts:
 \begin{eqnarray}
 P_T(\rm jet) \geq 150\ {\rm GeV}\,, ~~ P_T(\gamma)\geq 300\  {\rm GeV}\,,
 \end{eqnarray}
where $P_T$ denotes the transverse momentum of a particle or a jet. It can be seen that $\sigma(pp\to (\gamma\gamma)_H + {\rm jet} )$ in our model can be larger than $\sigma^{\rm SM}(pp\to \gamma\gamma + {\rm jet})$ after the kinematical cuts. For clarity, we also show the corresponding significance, which is defined by $S/\sqrt{B}$, in Fig.~\ref{fig:Hjet}(b), where we have fixed $y_F=-8$ and the different lines are associated with different luminosities. It can be found that with a luminosity of $60$ fb$^{-1}$, the significance can be above $4\sigma$ for $m_F< 900$ GeV. 

\section{Summary}

The minimal renormalizable gauge theory that provides a new charged gauge boson is a local $SU(2)$ gauge symmetry. Thus, it is of interest to study the model with $SU(2)_1\times SU(2)_2 \times U(1)_Y$ gauge symmetry.  The  $SU(2)_2$ gauge symmetry can be spontaneously broken by a $SU(2)_2$ doublet $H_2$. To avoid the complicated gauge anomaly cancellation, we consider an $SU(2)_2$ vector-like quark doublet $Q'$ so that  the charged gauge boson  $W'$ can decay into the vector-like quarks and SM quarks, and  the production cross section and decay branching ratios  of the heavy Higgs boson $H$ of $H_2$   can be enhanced.  

It is found that to satisfy the precision $\rho$-parameter measurement, the masses of new gauge bosons have to be heavier than 1.7 TeV; the upper limits of the current dilepton resonance experiments can give a strict bound on the new gauge coupling $g_2$; and the bound of $g_2$ from $\rho$-parameter becomes stronger when $g_2 > 2$. 

We add a scalar singlet $S$ in the model so that the $H$ boson can be produced via the gluon-gluon fusion channel due to  the mixing effect between $S$ and $H$.  It is found that the $W'$ gauge boson plays an important role in the $H$ to diphoton decay. As a result, the production cross section for $pp\to H \to \gamma\gamma$ can reach the upper limits of  the ATLAS and CMS experiments.   To illustrate the interesting collider signature, we study the process  $pp\to \gamma\gamma + {\rm jet}$ and  its significances with various luminosities. By taking proper values of Yukawa coupling $y_F$ and mass of vector-like quark, the significance can easily be over $4\sigma$.   
In addition, other possible signatures to illustrate the new physics effects in our model are  the production of vector-like quarks via $pp \to (\bar  T T, \bar B B)$, in which the dominant decay modes are $T \to t H$ and $B \to b H$. We found that the production cross section can reach  $0.4$ fb for $pp \to \bar Q' Q' \to \bar q q (\gamma \gamma)_H [\bar q q (gg)_H]$, which can be tested at the LHC.    
\\

\noindent{\bf Acknowledgments} \\

The work of CHC was  supported by the Ministry of Science and Technology of  Taiwan,
R.O.C., under grant  MOST-103-2112-M-006-004-MY3. 


\end{document}